# Dynamic density functional theory of multicomponent cellular membranes

L. G. Stanton,[1] T. Oppelstrup,[2] T. S. Carpenter,[2] H. I. Ingólfsson,[2] M. P. Surh,[2] F. C. Lightstone,[2] and J. N. Glosli[2]
[1]*Department of Mathematics and Statistics, San José State University, San José, California 95192, USA*
[2]*Physical and Life Sciences Directorate, Lawrence Livermore National Laboratory, Livermore, California 94550, USA*

We present a continuum model trained on molecular dynamics (MD) simulations for cellular membranes composed of an arbitrary number of lipid types. The model is constructed within the formalism of dynamic density functional theory and can be extended to include features such as the presence of proteins and membrane deformations. This framework represents a paradigm shift by enabling simulations that can access length scales on the order of microns and time scales on the order of seconds, all while maintaining near fidelity to the underlying MD models. These length and time scales are significant for accessing biological processes associated with signaling pathways within cells. Membrane interactions with RAS, a protein implicated in roughly 30% of human cancers, are considered as an application. Simulation results are presented and verified with MD simulations, and implications of this new capability are discussed.

## I. INTRODUCTION

Cellular membranes are vital to a myriad of biological functions, as they not only control the structure and permeability of a cell but are also involved in processes such as cell adhesion, ion conductivity and cell signaling through the presence of membrane proteins. As such, the modeling of cellular membranes and their interaction with membrane proteins is of great interest to predict membrane dynamics at scales that are inaccessible by experiments. Unfortunately, as we discuss below, there is a significant gap between the resolution of most experiments and the scales that atomically-resolved simulations can achieve. Worse, many biological phenomena of interest, such as the migration and signaling of membrane proteins, exist in this elusive regime.

The primary simulation methods for modeling cellular membranes are molecular dynamics (MD) and Monte Carlo (MC) [1,2], in which either individual atoms or biologically-relevant clusters of atoms are evolved according to an effective force field [3–7]. As with any MD simulation, there will be limitations on the accessible time and length scales, which prevents MD from probing many biological problems. At current capabilities, meaningful MD simulations of membranes can achieve length scales of order $10^4$ nm$^2$ and time scales of order 10 $\mu$s [8].

Continuum models are sometimes used to achieve larger scales but usually at the sacrifice of both accuracy and generality. Phase field models are a common continuum approach [9–12], which have been extended using stochastic Saffman-Delbrück hydrodynamics [13–15]. The primary challenge of these continuum models is that they rely on phenomenological constructions of the free energy, which must be fit to either experimental or MD simulation data. Furthermore, the lipid bilayer of biological membranes typically consists of hundreds of different lipid types [16–18], and a detailed description should resolve all common species. Unfortunately, phase field models are prone to an explosion of fitting parameters with an increasing number of species, and consequently, most phase field models of cellular membranes only include 2–3 species. To mitigate these challenges, multiscale methods are paramount in accessing these larger scales with continuum models while still maintaining some level of atomic-scale accuracy.

Classical density functional theory (DFT) is a formalism that not only connects macroscopic quantities with microscopic degrees of freedom without the need for empirical parameters, but it also naturally generalizes to multicomponent systems with an arbitrary number of species. Furthermore, dynamic DFT (DDFT), which is a nonequilibrium extension of classical DFT [19,20], thus presents an alternate approach to modeling cellular membranes that are fundamentally both multicomponent and dynamic. Like many other continuum models used for describing cellular membranes, DDFT evolves densities in accordance with an appropriate free energy functional

$$\mathscr{F}[\mathbf{n}] = \int_V f(\mathbf{n})\, d\mathbf{r}, \qquad (1)$$

where $f(\mathbf{n})$ is the free energy density, $\mathbf{n} = \{n_i(\mathbf{r}, t)\}$ is a set of densities, and $V$ is the domain of interest. DDFT is a growing field in modern statistical physics [21], with applications in both the physical sciences, such as polymers [22], colloids [23], granular media [24] and oxidation [25], as well as the life sciences, such as protein absorption [26], tumor growth [27], and epidemiology [28]. There have also been recent advances to extend the field to hydrodynamic DFT for applications like plasmas in which inertial and viscoelastic effects are relevant [29–31]. The primary goal of this paper is to provide the theoretical framework for modeling cellular membranes using DDFT, which has already been shown to be successful in practice [32–34].

The document is organized as follows. The mathematical model is described in full in Sec. II, with simulation results of this model being presented in Sec. III. Finally, conclusions and a discussion of these results are discussed in Sec. IV.

## II. MODEL FORMULATION

In this section, we derive the governing equations of motion, which primarily relies on the construction of an appropriate free energy functional of the system, as detailed in Sec. II A. From this free energy, the corresponding dynamics can be inferred for the membrane lipids as well as any macromolecules embedded in the membrane such as proteins, as detailed in Sec. II B.

### A. Free Energy Functional

The total free energy of the membrane will thus depend on lipid-lipid, lipid-protein, and protein-protein interactions and be decomposed in terms of these interactions as $f = f_{\ell\ell} + f_{\ell p} + f_{pp}$, respectively. To determine an approximate form for the lipid-lipid contribution, we can consider a membrane being composed of $N$ species of lipid headgroups, where species are additionally distinguished if they are on the inner or outer leaflet of the membrane, and each species density $n_i = n_i(x, y, t)$. Unlike phase field models, which approximate the free energy density through gradient expansions of the density (or a relevant order parameter), DDFT models capture these interactions through a correlation expansion, which formally connect to the microscopic statistics of the system. Within the Ramakrishnan-Yussouff (RY) approximation [35], which truncates this expansion to second-order and assumes spatial isotropy, we can express the free energy density as

$$f_{\ell\ell} \approx k_B T \sum_{i=1}^{N} \left( n_i [\log(\Lambda^2 n_i) - 1] - \frac{1}{2} \sum_{j=1}^{N} \int_V \Delta n_i(\mathbf{r}) c_{ij}(\mathbf{r} - \mathbf{r}') \Delta n_j(\mathbf{r}') \, d\mathbf{r}' \right). \quad (2)$$

Here, $\Delta n_i(\mathbf{r}) = n_i(\mathbf{r}) - \bar{n}_i$ is the density fluctuation about its mean $\bar{n}_i$, $k_B$ is the Boltzmann constant, $T$ is the temperature and $\Lambda$ is the de Broglie wavelength, which will cancel out upon taking the gradients within the evolution equations. Finally, $c_{ij}(\mathbf{r})$ is the direct correlation function (DCF) between lipid species $i$ and $j$, which can be calculated using atomically-resolved simulations such as MD. For example, two-dimensional radial distribution functions (RDFs) $g_{ij}(r)$ can be calculated between the lipid headgroups, and the corresponding DCFs can be constructed by inverting the Ornstein-Zernike (OZ) relations [36] given by

$$h_{ij}(r) = c_{ij}(r) + \sum_{i'=1}^{N} \bar{n}_{i'} h_{ii'}(\mathbf{r}) * c_{i'j}(\mathbf{r}), \quad (3)$$

$$f(\mathbf{r}) * g(\mathbf{r}) \equiv \int f(\mathbf{r}')g(\mathbf{r} - \mathbf{r}')d\mathbf{r}', \quad (4)$$

where the *total* correlation function is simply defined as $h_{ij}(r) \equiv g_{ij}(r) - 1$, and the operation $(*)$ denotes a convolution. Details of our $g_{ij}(r)$ calculation are discussed below.

Due to the low collisionality of proteins on the membrane, a continuum description is inappropriate. Instead, each protein can be represented as one or more "beads" with position $\mathbf{r}_k(t)$. Given $P$ beads, the lipid-protein energy density can be expressed as

$$f_{\ell p} = \sum_{i=1}^{N} \sum_{k=1}^{P} n_i(\mathbf{r}) u_{ik}(\mathbf{r} - \mathbf{r}_k), \quad (5)$$

where $u_{ik}(\mathbf{r})$ is the potential of mean force (PMF) between the $k$th protein and the $i$th lipid species. Note that some of these PMFs are potentially zero in cases where the protein interacts with only one leaflet. As with the lipid-lipid interactions, the lipid-protein PMF can be calculated consistently with the RY approximation as

$$u_{ik}(r) = k_B T [h_{ik}(r) - c_{ik}(r) - \log(g_{ik}(r))]. \quad (6)$$

Note that Eq. (6) is identical to the hypernetted chain (HNC) closure relation [37]. While the lipid-protein RDFs can be calculated directly from atomically-resolved MD or MC simulations, the lipid-protein DCFs must be constructed using Eq. (3) along with the additional OZ equations that include the contributions from the lipid-protein interactions as well:

$$h_{ik}(r) = c_{ik}(r) + \sum_{j=1}^{N} \bar{n}_j h_{ij}(\mathbf{r}) * c_{jk}(\mathbf{r}). \quad (7)$$

Similar to the lipid-protein interactions, the protein-protein energy density can be expressed as

$$f_{pp} = \frac{1}{2} \sum_{k=1}^{P} \sum_{k'=1}^{P} \delta(\mathbf{r} - \mathbf{r}_k) u_{kk'}(\mathbf{r} - \mathbf{r}_{k'}), \quad (8)$$

where the continuous density fields have been replaced with a distribution of point beads expressed here using Dirac delta functions $\delta(\mathbf{r})$, and each sum is now only over the bead indices $k$. Note that despite $f_{pp}$ being a free energy density, we have omitted contributions from the configurational entropy, as they are expected to be negligible when compared to the lipid-protein interactions. Understanding the interactions between proteins has led to a vast field of study, and even atomistic simulations have proven to be too computationally expensive to fully parametrize them. As such, a complete treatment of these interactions is beyond the scope of this work, and we simply rely on a phenomenological approach by assuming that the interactions are composed of an attractive and a repulsive component. An example of such an interaction is given by the Mie (or Kihara) potential, which is a generalization of the Lennard-Jones potential for nonspherical molecules [38,39]. This potential can be further generalized by shifting the radial dependence as $r \rightarrow |\mathbf{r} - \mathbf{r}_0|$, which introduces an excluded volume to account for the finite radius ($r_0$) of a given bead.

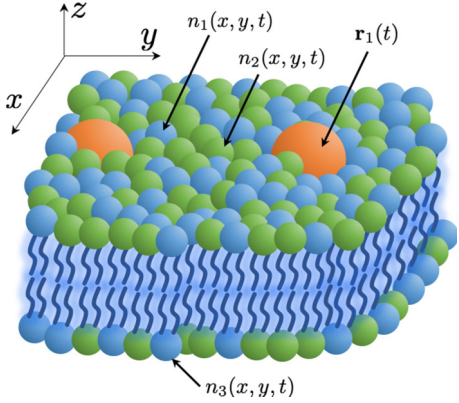

FIG. 1. Schematic of a lipid bilayer with proteins embedded in the membrane. Examples of different lipid densities are indicated with different colors on each leaflet. In the full simulations, 7 lipid densities in the inner leaflet and 6 lipid densities in the outer leaflet were considered with 10s of proteins attached to the inner leaflet.

Equations (2), (5), and (8) thus form the leading order components to the free energy density of the lipid-membrane system in our model. An example schematic of the mathematical domain and relevant variables for a lipid bilayer with several embedded proteins is shown in Fig. 1.

### B. Evolution Equations

Once the free energy functional is constructed, the evolution of the lipid density fields and protein beads can be determined. Each lipid density field, which is an inherently conserved quantity in the absence of chemical reactions, can be described using a continuity equation

$$\frac{\partial n_i}{\partial t} + \nabla \cdot (n_i \mathbf{v}_i) = 0, \quad (9)$$

where $\mathbf{v}_i$ is the associated velocity field. Rather than evolve the velocity fields as well, which will begin a hierarchy of evolution equations for each velocity moment, we can instead close this hierarchy by exploiting the disparate timescales associated with the lipid dynamics. Due to the extremely small Reynolds number at this scale ($Re \sim 10^{-6}$), the velocity fields will rapidly relax when any force is applied, and thus inertial velocity fields can be neglected when compared to the much larger viscous forces at this scale. We can thus separate each velocity into dissipating and fluctuating components as $\mathbf{v}_i = \mathbf{v}_i^d + \mathbf{v}_i^f$, where the first term will be approximated in the Stokes limit as $\mathbf{v}_i^d \approx \beta D_i \mathbf{F}_i^d$, and the second term will be approximated with a stochastic model $\mathbf{v}_i^f \approx \boldsymbol{\xi}_i$. Here, $\boldsymbol{\xi}_i(\mathbf{r}, t)$ is a vector Wiener process with the statistical properties $\langle \boldsymbol{\xi}_i(t) \rangle = \mathbf{0}$ and $\langle \xi_{i,j}(t) \xi_{i,k}(t') \rangle = (2k_B T/m_i) \delta_{jk} \delta(t - t')$ for each vector component, with $\delta_{ij}$ being the Kronecker delta and $m_i$ being the lipid mass. Lastly, $\mathbf{F}_i^d$ is the lateral drag force on the lipids, $\beta = 1/(k_B T)$ is the thermodynamic beta, and $D_i$ is the diffusivity of lipid type $i$, which can be calculated from MD. The drag force can be expressed in terms of the chemical potential (and thus the free energy) as

$$\mathbf{F}_i^d = -\nabla \mu_i = -\nabla \left( \frac{\delta \mathscr{F}}{\delta n_i} \right), \quad (10)$$

where the $\delta$ denotes a variational derivative.

Combining these expressions for the velocity field components into Eq. (9) thus gives the DDFT equation

$$\frac{\partial n_i}{\partial t} = \nabla \cdot \left( \beta D_i n_i \nabla \left( \frac{\delta \mathscr{F}}{\delta n_i} \right) + n_i \boldsymbol{\xi}_i \right). \quad (11)$$

It should be noted that the final term, which encapsulates the fluctuations, will maintain conservation of the lipids by construction. These fluctuation effects are more commonly included as an additive term, which then requires a modification to the correlation properties to maintain conservation of the density [40]. Evaluating the variational derivative of the free energy with respect to $n_i$ yields

$$\frac{\delta \mathscr{F}}{\delta n_i} = k_B T \left( \log(\Lambda^2 n_i) - \sum_{j=1}^{N} \Delta n_j * c_{ij} \right) + \sum_{k=1}^{P} u_{ik}(\mathbf{r} - \mathbf{r}_k). \quad (12)$$

Finally, the coordinates associated with the protein beads will evolve in accordance with the energetic contributions $f_{\ell p}$ and $f_{pp}$. These equations of motion can be generalized to include the missing degrees of freedom associated with fluctuations in the cytoplasm through a set of Langevin equations:

$$\frac{d^2 \mathbf{r}_k}{dt^2} = \frac{1}{m_k} \mathbf{F}_k - \gamma_k \frac{d \mathbf{r}_k}{dt} + \boldsymbol{\theta}_k(t), \quad (13)$$

$$\mathbf{F}_k = -\nabla \sum_{k' \neq k}^{P} u_{kk'}(\mathbf{r} - \mathbf{r}_{k'}) \bigg|_{\mathbf{r}=\mathbf{r}_k} - \nabla \sum_{i=1}^{N} u_{ki}(\mathbf{r}) * n_i(\mathbf{r}) \bigg|_{\mathbf{r}=\mathbf{r}_k}, \quad (14)$$

where $\boldsymbol{\theta}_k$, like $\boldsymbol{\xi}_i$, is a vector Wiener process with zero mean and the delta-correlated variance $\langle \theta_{k,i}(t) \theta_{k,j}(t') \rangle = 2 k_B T \gamma_i \delta_{ij} \delta(t - t')$. The relaxation rate is obtained from the fluctuation-dissipation theorem as $\gamma_k = k_B T / m_k \mathcal{D}_k$, where $\mathcal{D}_k$ is the diffusivity of the protein, which can be calculated from MD through the appropriate Green-Kubo relation [41,42].

### C. Membrane Deformations

It should be noted that an obvious extension of the model is to allow for membrane deformations, which can be done through adding the so-called Helfrich Hamiltonian to the free energy [43]. For a single component system, the associated Helfrich energy density is given by

$$f_H = \frac{\kappa}{2} (\nabla^2 h - R^{-1})^2, \quad (15)$$

where $\kappa$ is the bulk modulus, $z = h(x, y, t)$ is the mean deformation field of the membrane, and $R^{-1}$ is the spontaneous curvature of the membrane that represents the equilibrium curvature measured from cylindrical micelles (typical spontaneous curvatures can range up to $|R^{-1}| \sim 20\,\text{nm}^{-1}$) [9,10]. Note that $\nabla^2 h$ is twice the linearized mean curvature of the membrane and is thus the appropriate measure of deviation from the spontaneous curvature. Unfortunately, it is *currently*

TABLE I. Definitions of lipid types used in the DDFT simulations along with their corresponding concentrations and self-diffusion coefficients ($D_i$) within the inner and outer leaflets. In the Martini CG representation, 4 to 1 heavy atoms per CG bead, each lipid tail represent more than one physical lipid tail length [e.g., a P CG tail maps to both a stearoyl (C18) and a palmitoyl (C16)].

| Label | Representative Name | Outer Fraction | Inner Fraction | Outer $D_i$ (μm²/s) | Inner $D_i$ (μm²/s) |
|---|---|---|---|---|---|
| CHOL | cholesterol | 31.3% | 28.0% | 43 | 43 |
| POPC | 1-palmitoyl-2-oleoyl-sn-glycero-3-phosphocholine | 24.3% | 13.9% | 36 | 46 |
| PAPC | 1-palmitoyl-2-arachidonoyl-sn-glycero-3-phosphocholine | 12.1% | 9.7% | 36 | 44 |
| POPE | 1-palmitoyl-2-oleoyl-sn-glycero-3-phosphoethanolamine | 2.1% | 5.4% | 31 | 39 |
| DIPE | 1,2-dilinoleoyl-sn-glycero-3-phosphoethanolamine | 6.1% | 16.1% | 34 | 45 |
| DPSM | N-stearoyl-D-erythro-sphingosylphosphorylcholine | 24.2% | 10.8% | 35 | 45 |
| PAPS | 1-palmitoyl-2-arachidonoyl-sn-glycero-3-phosphatidylserine | --- | 16.1% | --- | 49 |

*unknown* how to generalize Eq. (15) to multicomponent systems [44]. Two popular suggestions are given by the *local* summation over curvatures

$$f_{H,\ell} = \frac{\kappa}{2} \sum_{i=1}^{N} c_i (\nabla^2 h - R_i^{-1})^2 \qquad (16)$$

and the *global* summation over curvatures

$$f_{H,g} = \frac{\kappa}{2} \left( \nabla^2 h - \sum_{i=1}^{N} c_i R_i^{-1} \right)^2. \qquad (17)$$

In each representation, the lipid concentrations are given by $c_i = n_i / \sum_j n_j$, and $R_i$ is the spontaneous curvature of a given lipid species. The situation is further complicated when accounting for a composition-dependent bulk modulus as well as the inner and outer leaflets of the lipid bilayer, which necessitates a leaflet dependence of each spontaneous curvature and the potential for multiple deformation fields.

For these reasons and the fact that the underlying MD simulations in this study only exhibit relative membrane deformations of a few percent, we will ignore curvature affects and assume the cell membrane to be a two-dimensional surface. Additionally, supported bilayers, commonly used for membrane experiments, provide a relevant example of membranes that can reach length scales of many microns while remaining essentially flat. For systems in which deformations are appreciable, further research is required for the necessary multi-component generalization.

### III. SIMULATION RESULTS

We turn our general formalism to the particular application of modeling the lipid interaction and aggregation of RAS proteins on the inner leaflet of a plasma membrane (PM) mimic. In its mutated state, RAS has been implicated in ∼30% of human cancers (95% of pancreatic cancers, 45% colorectal cancers, *etc.*) [45,46], and a better understanding of RAS-lipid interactions and aggregate formation in oncogenic signaling pathways is sorely needed for this often-labeled "undruggable target" [47].

To model this system of interest, we take KRAS4b proteins (a frequently mutated RAS isoform in human cancers [48,49]) inserted on the inner leaflet of a PM mimic. In particular, this lipid bilayer consists of an asymmetric seven-lipid mixture tuned to represent average PM bulk lipid properties, based on the eight-component mixture defined in [50] without the inner leaflet PIP2 lipid. The outer leaflet contains POPC, PAPC, POPE, DIPE, DPSM, and CHOL with more of the saturated/ordered lipids and cholesterol, whereas the inner leaflet also contains PAPS (a negatively charged lipid), less cholesterol, and more unsaturated lipids [50]. A table of the lipid definitions, their concentrations and their diffusivities on the inner and outer leaflets are given in Table I.

The primary inputs to the DDFT model are then the 91 lipid-lipid RDFs and seven inner leaflet lipid-protein RDFs obtained from MD simulations, which are in turn converted to DCFs and PMFs using Eqs. (3), (6), and (7), and the 13 self-diffusion coefficients for the lipids on each leaflet. To obtain these inputs, a Martini MD simulation [6] of a 30 × 30 nm² bilayer with one protein was run for 32 μs using similar MD parameters, setup, and input parameter extraction as in [33]. The left panel of Fig. 2 shows a subset of these DCFs and PMFs for the model.

As a test of model validity, we seek to compare the RDFs obtained from MD with RDFs predicted from the DDFT simulations. To compute lipid-protein RDFs using the DDFT model, a protein was fixed in space, while time-averaged densities of the lipids were calculated as a function of distance from the protein. To introduce a similar point source for the computation of the lipid-lipid RDFs, an external density with a narrow Gaussian profile corresponding to the area of one lipid molecule was added to the system. This source was then assigned the appropriate DCF as time-averaged densities were once again calculated. Examples of true (MD) and predicted (DDFT) RDFs are shown in Figs. 3 and 4, where strong agreement can be seen. Absolute errors between lipid-protein RDFs were observed to be small, and while larger disagreements for lipid-lipid RDFs were observed at small $r$, neighbor counts in this region revealed relative errors that were at most ∼2%. It should also be noted that these larger errors appear to be due to both the softened interactions in the DDFT model as well as the poor sampling of MD at that small-$r$ scale. Similar agreement was found in the remaining cases but were to extensive to present in this work.

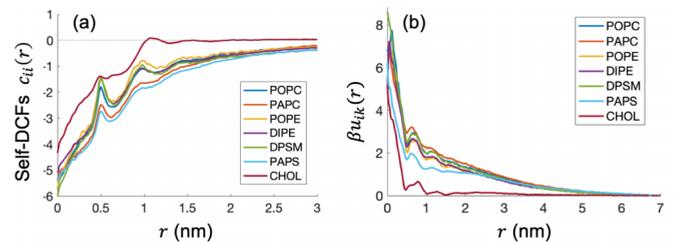

FIG. 2. (a) Self-DCFs $c_{ii}(r)$ computed using RDFs from MD and the OZ Eq. (3). (b) Dimensionless protein-lipid potentials $\beta u_{ik}$ computed using lipid-protein RDFs MD, OZ Eqs. (3) and (7), and the HNC closure [Eq. (6)].

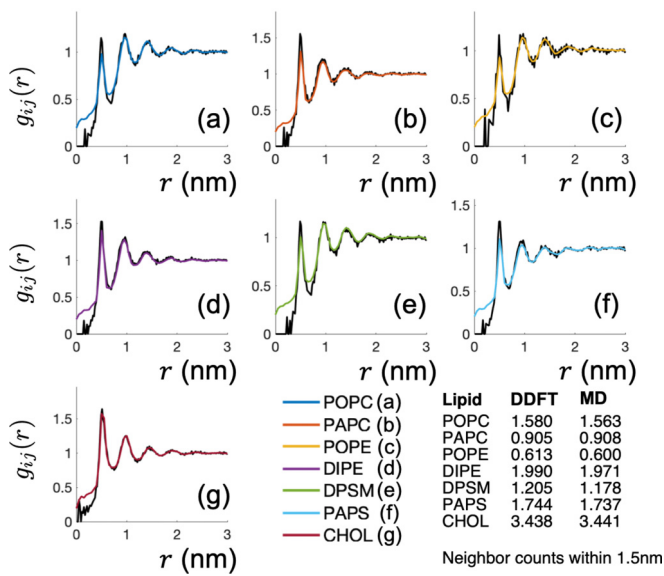

FIG. 3. Lipid-lipid RDFs for PAPS computed using the DDFT (colors same as in Fig. 2) compared to MD (black) used to parameterize the model. Though the largest disagreement between RDFs is seen at small $r$, neighbor counts within a radius of $r = 1.5$ nm show that relative errors for this measure are at most ∼2%.

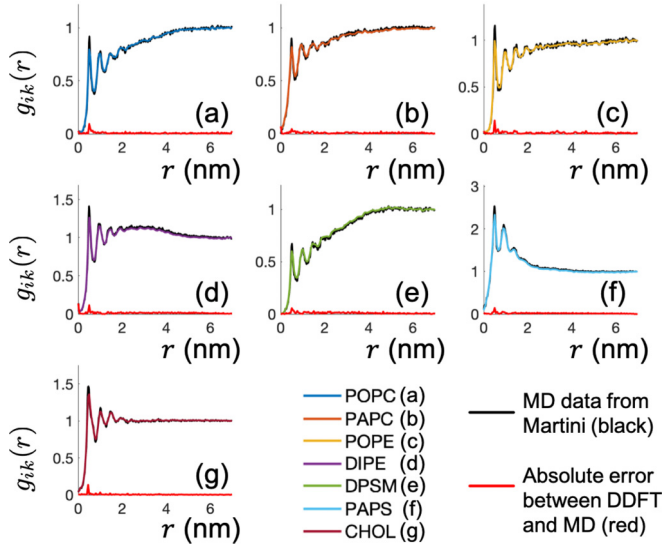

FIG. 4. Comparison of lipid-protein RDFs from DDFT (color curves) and MD (black). The absolute error between the two methods is shown in red.

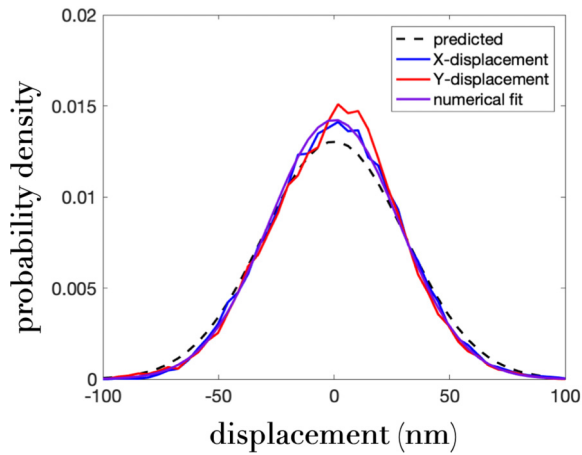

FIG. 5. Probability distributions of the displacement of a single protein. Displacements of the $x$ and $y$ components, which are sampled $2\,\mu$s apart, are the red and blue curves, respectively. The distribution associated with $\mathcal{D}_k = 23.4\ \mu m^2/s$, which is the input diffusivity to the Langevin model, is the dashed curve. A numerical fit to the simulation data is shown in purple, which corresponds to a diffusivity of $\mathcal{D}_k = 19.6\ \mu m^2/s$. As expected, we see that the simulated displacement is reduced due to lipid-protein interactions.

Furthermore, we measure the diffusivity of a single protein on the inner leaflet. The input diffusivity used in the Langevin model for the protein is set to $\mathcal{D}_k = 23.4\ \mu m^2/s$; however, the interactions with the lipids will alter this value. Probability distributions of the protein displacement are shown in Fig. 5, where the $x$ and $y$ displacements are sampled $2\,\mu$s apart. Additionally, the distribution associated with $\mathcal{D}_k = 23.4\ \mu m^2/s$ is compared with a numerical fit to the data. As expected, we see that the actual displacement corresponds to a slightly slower diffusion coefficient ($19.6\ \mu m^2/s$), due to lipid-protein interactions. It should be noted that the diffusion rates obtained at the MD scale are highly susceptible to finite size effects from the simulation [51–53]. Additionally, diffusion will be artificially high (roughly a factor of 4) due to the smoother potentials associated with the course-grained Martini force field [6]. Rather than make these corrections to the diffusion rates, we were careful in enforcing all MD simulations to be roughly the same size, so that all of our DDFT parameters (*e.g.*, rates and populations) would be self-consistent. For a more thorough discussion of how the RAS diffusion rates were calculated, see Sec. 2.7 of the Supplemental Material in Ref. [33].

To demonstrate this capability, we simulate a region of the cellular membrane with 20 RAS proteins, shown in Fig. 6. In particular, the RAS proteins can be seen to form in clusters despite the simplistic protein-protein interactions, which are taken to be purely repulsive in these simulations. This clustering is therefore due to an implicit attraction induced by the lipid-protein and lipid-lipid interactions and is believed to be a lipid-dependent mechanism for the colocalization of RAS, which amplifies downstream signaling, making it a target for modulating excessively active oncogenic RAS [33].

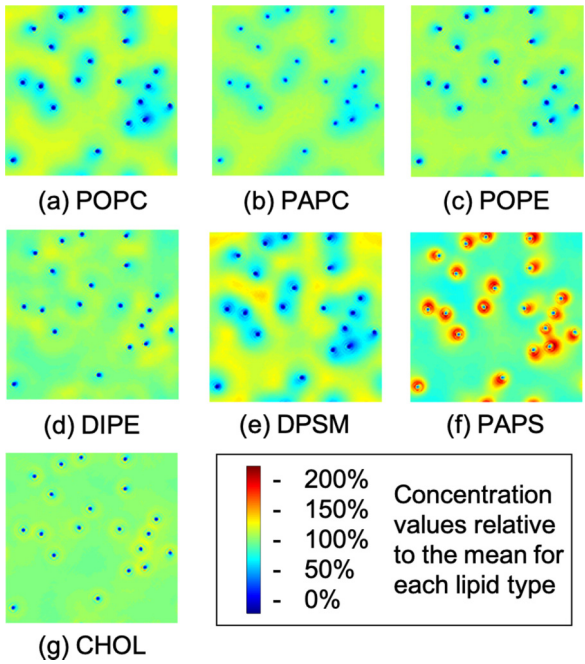

FIG. 6. Concentrations $n_i(\mathbf{r}, t)/\bar{n}_i$ for inner leaflet lipids in $30 \times 30$ nm$^2$ simulation with 20 RAS proteins. In each plot, green (100%) indicates the average density for that lipid.

We can further investigate the possibility of protein aggregation through perturbations of the lipid-protein interactions. In particular, we choose to perturb the PAPS-RAS interaction by augmenting the potential with a term of the form $-\alpha T r e^{-1.5r}$, where $r$ is the lipid-protein distance, and $\alpha$ is a positive, dimensionless parameter quantifying the magnitude of the perturbation. This added contribution effectively models proteins with different levels of affinity for PAPS. Examples of the perturbed and unperturbed PAPS-RAS interactions are shown in the top panel of Fig. 7. The simulations with 20 RAS proteins in the $30 \times 30$ nm$^2$ domain were repeated with the modified PAPS-RAS interaction, and as can be seen in Fig. 8, the system exhibits a transition towards protein aggregation as the perturbation parameter $\alpha$ is increased. The bottom panel of Fig. 7 shows the RAS-RAS neighbor density distribution for different $\alpha$ and confirms the visually observed aggregation of Fig. 8.

The transition to protein aggregation can be quantified by calculating the number of protein neighbors a given protein has within a 4 nm radius (roughly the first neighbor shell) and taking the ensemble average over the simulation. Alternatively, we can calculate the height of first peak in the neighboring protein density distribution. As shown in Fig. 9, both the protein neighbor count (left panel) and the neighbor density peak (right panel) experience a sharp increase as the perturbation parameter $\alpha$ is varied, with an inflection point observed at roughly $\alpha \sim 10$, which corresponds to island formation and aggregation of proteins as seen in Fig. 8. While this transition appears to be continuous, it should be noted that the system is finite and only includes 20 proteins. We expect that increasing the domain would lead to more discontinuous changes in these neighbor counts as $\alpha$ is varied, consistent with a true phase transition.

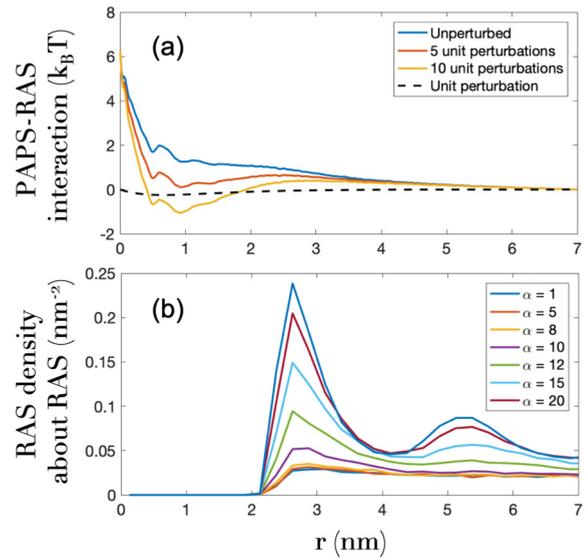

FIG. 7. (a) Effect of the perturbation to the PAPS-RAS interaction. The increase of the parameter $\alpha$ enhances the attraction between the RAS proteins and PAPS lipids. (b) RAS neighbor density distributions about RAS as a function $\alpha$ in units of lipids/nm$^2$.

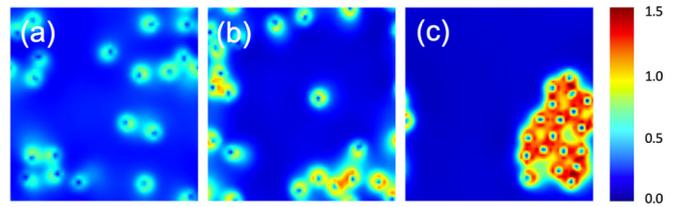

FIG. 8. Snapshots of PAPS density fields for various modifications to the PAPS-RAF interaction. The perturbation parameters shown are (a) $\alpha = 5$, (b) $\alpha = 10$, and (c) $\alpha = 15$.

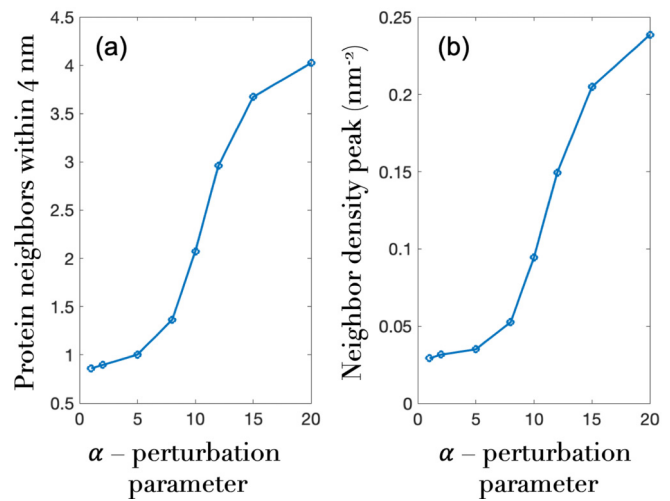

FIG. 9. Quantification of the transition towards RAS aggregation. (a) Average number of RAS-RAS neighbors within 4 nm. (b) Peak of the protein neighbor density profile in units of proteins/nm$^2$. Both curves show a steep transition near $\alpha \sim 10$.

## IV. CONCLUSIONS

In summary, we have shown that DDFT provides an ideal framework for modeling multicomponent, cellular membranes as a continuum by incorporating the underlying physics at the molecular scale in a rigorous and self-consistent way. Once DCFs and PMFs are constructed from MD (or MC) simulations, the DDFT model can be used to access time and length scales many orders of magnitude beyond those of MD simulations with a relatively low sacrifice to accuracy.

An application of this model is considered in which the aggregation of RAS proteins is explored, a potential mechanism for the signaling pathway of cancer growth. Due to the computational efficiency of the DDFT model, the parameter space associated with the empirical protein interactions could be explored with relative ease, and a phase transition associated with the strength of the PAPS-RAS interaction was observed.

In future studies, more accurate protein interactions will be included in addition to effects of membrane deformations. Furthermore, the model will be extended to account for systems in which the RDFs evolve in an appreciable way, necessitating a feedback loop between the continuum and molecular scales.

## ACKNOWLEDGMENTS

This work has been supported in part by the Joint Design of Advanced Computing Solutions for Cancer (JDACS4C) program established by the US Department of Energy (DOE) and the National Cancer Institute (NCI) of the National Institutes of Health (NIH). We thank the entire JDACS4C Pilot 2 team, particularly the Pilot 2 leads Fred H. Streitz and Dwight V. Nissley, for their support and helpful discussion. For computing time, we thank Livermore Computing (LC) and Livermore Institutional Grand Challenge. This work was performed under the auspices of the US DOE by Lawrence Livermore National Laboratory under Contract No. DE-AC52-07NA27344 and used resources of the Oak Ridge Leadership Computing Facility, which is a DOE Office of Science User Facility supported under Contract No. DE-AC05-00OR22725. The document number is LLNL-JRNL-829309.